\documentstyle[12pt]{article}
\begin{document}
\begin{titlepage}
\mbox{ }
\rightline{UCT-TP-208/94}
\rightline{April 1994}
\vspace{1.5cm}
\begin{center}
\begin{Large}
{\bf  Comment on current correlators in QCD at finite temperature}

\end{Large}

\vspace{2cm}

{\large {\bf C. A. Dominguez}}

Institute of Theoretical Physics and Astrophysics, University of
Cape Town, Rondebosch 7700, South Africa\\

\vspace{3mm}

and \\

\vspace{3mm}

{\large {\bf M. Loewe}}\\[0.5cm]
Facultad de Fisica, Pontificia Universidad Cat\'{o}lica
de Chile, Santiago, Chile\\
\end{center}

\vspace{1.6cm}

\begin{abstract}
\noindent
We address some criticisms by Eletsky and Ioffe on the extension of
QCD sum rules to finite temperature. We argue that this extension is
possible, provided the Operator Product Expansion and QCD-hadron
duality remain valid at non-zero temperature. We discuss evidence in
support of this from QCD, and from the exactly solvable  two-
dimensional
sigma model O(N) in the large N limit, and the Schwinger model.
\end{abstract}
\end{titlepage}

\setlength{\baselineskip}{1.5\baselineskip}
\noindent
Some time ago Bochkarev and Shaposnikov \cite{ref1} proposed an
extension of
the QCD sum rule program to non-zero temperature, and made an
application
to the two-point function involving the vector current. This
application was reconsidered in \cite{ref2}.
Later on we discussed the axial-vector channel \cite{ref3} and the
nucleon channel \cite{ref4} using
Finite Energy QCD sum rules (FESR). The results from these analyses
indicate
a substantial rearrangement of the hadronic spectrum with increasing
temperature, and hint on the existence of a deconfining phase
transition.
This was later confirmed in \cite{ref5}, where a formalism valid even
for
T near the critical temperature was used. QCD sum rules \cite{ref6}
are
based on the Operator Product Expansion (OPE) of current correlators
at
short distances, suitably extended to include non-perturbative
effects. The
latter are parametrized in terms of a set of vacuum expectation
values of
the quark and gluon fields entering the basic QCD Lagrangian. Contact
with
the hadronic world of large distances is achieved by invoking the
notion
of QCD-hadron duality. The values of the vacuum condensates cannot be
calculated analytically from first principles, as this would be
tantamount
to solving QCD exactly. Instead, they are extracted from certain
channels
where experimental information is available, e.g. $e^{+}e^{-}$
annihilation,
and $\tau$ decays \cite{ref7}. It is also possible, in principle,
to estimate them numerically from lattice QCD.
An extension of this formalism of QCD sum
rules to finite temperature entails the assumptions that (a) the OPE
continues to be valid, except that now the vacuum condensates will
develop
an ({\it a-priori}) unknown temperature dependence, and (b) the
notion of
QCD-hadron duality also remains valid. In analogy with the situation
at T=0,
the thermal behaviour of the vacuum condensates is not calculable
analytically from first
principles. Some model or approximation must be invoked, e.g. the
dilute
pion gas approximation, lattice QCD, etc..
The quark, the gluon, and the four-quark condensates at
finite temperature have thus been estimated in the literature
\cite{ref8}-
\cite{ref10}.

In a recent paper, Eletsky and Ioffe \cite{ref11} have criticized
the QCD sum rule program proposed in \cite{ref1}, and developed
in \cite{ref2}-\cite{ref5}.
In this note we wish to address this criticism, and hopefuly
clarify the issue. We will argue that, provided the  OPE
and  QCD-hadron duality remain valid at finite temperature, the
approach of \cite{ref1} -\cite{ref5} is basically correct. We
illustrate our argument with the current correlator involving the
axial-vector current, although it can be trivially generalized to any
local current operator. In addition, we provide further supportive
evidence from two exactly solvable field theory models: the two-
dimensional
O(N) sigma model in the large N limit, and the two-dimensional
Schwinger
model.

The basic object to be considered is the retarded (advanced) two-point
function after appropriate Gibbs averaging
\begin{equation}
\Pi \; (q,T) = i \int d^{4}x \;\exp (i q x) \; \theta(x_{0})
 << [J(x), J^{\dag}(0)] >> \; ,
\end{equation}
where
\begin{equation}
<< A \cdot B >> = \sum_{n}\; \exp (-E_{n}/T)\; \langle n|A \cdot B|n
\rangle
\;/Tr (\exp (-H/T)) \; ,
\end{equation}
and $\mid n>$ is a complete set of eigenstates of the (QCD)
Hamiltonian.
The OPE of $\Pi (q,T)$ is formally written as
\begin{equation}
\Pi \; (q,T) = C_{I} << I >> + \sum_{r} C_{r} (q) << {\cal{O}}_{r} >>
\; ,
\end{equation}
where the Wilson coefficients $C_{r} (q)$ depend on the Lorentz
indices
and quantum numbers of the external current $J(x)$, and also of the
local
gauge-invariant operators ${\cal O}_{r}$ built from the quark and
gluon
fields of QCD. The $C_{r} (q)$ could also depend on the
temperature, but since this
is not essential for the argument we shall ignore this dependence.
The unit
operator $I$ in Eq.(3) represents the purely perturbative piece. The
OPE  is assumed valid, even in the presence of non-perturbative
effects,
for $q^{2} < 0$ (spacelike), and  $\mid q^{2} \mid \gg$ $\Lambda_{QCD}
^{2}$. In principle, all Wilson coefficients are calculable in
perturbative QCD to any desired order in the strong coupling
constant. In
the sequel we shall work at leading (one loop) order for simplicity.
The
non-perturbative effects are then buried in the vacuum condensates.
Since
these have dimensions, the associated Wilson coefficients fall off as
inverse powers of $Q^{2} = - q^{2}$.

For instance, if the current $J(x)$ in
Eq.(1) is identified with the axial-vector current
$ A_{\mu} (x) = : \bar{u}(x) \gamma_{\mu} \gamma_{5} d(x) :$, then
with
\begin{equation}
\Pi_{\mu \nu}(q,T) = - g_{\mu \nu} \Pi_{1} (q, T) + q_{\mu} q_{\nu}
\Pi_{0} (q,T)
\; ,
\end{equation}
one easily finds  at T=0 \cite{ref6}
\begin{equation}
8 \pi^{2} \Pi_{0} (q, T=0) = - \ln \frac{Q^{2}}{\mu^{2}} +
\frac{C_{4} \langle 0_{4} \rangle}{Q^{4}} + \frac{C_{6} \langle 0_{6}
\rangle}{Q^{6}}
+ \cdots \; ,
\end{equation}
where $\mu$ is a renormalization scale, and e.g. the leading vacuum
condensate is given by
\begin{equation}
C_{4} \langle 0_{4} \rangle = \frac{\pi}{3} \langle \alpha_{s} \;
G^{2} \rangle - 8 \pi^{2} \bar{m}_{q} \langle \bar{q} q \rangle \; ,
\end{equation}
with $\bar {m}_{q} = (m_{u} + m_{d})/2$, and $<\bar{q} q> = <\bar{u}
u>
\simeq <\bar{d} d>$. The function $\Pi_{0}(q)$, Eq. (5), satisfies
a dispersion relation
\begin{equation}
\Pi_{0} (Q^{2}) = \frac{1}{\pi} \int ds \; \frac{\mbox{Im} \; \Pi_{0}
(s)}{s + Q^{2}} \; ,
\end{equation}
defined in this case up to one subtraction constant, which can be
disposed
of by e.g. taking the first derivative with respect to $Q^{2}$ in
Eq.(7).
The notion of QCD-hadron duality is implemented by calculating the
left hand
side of Eq.(7) in QCD through the OPE, and parametrizing the spectral
function
entering the right hand side in terms of hadronic resonances, followed
by a hadronic continuum modelled by perturbative QCD. In this fashion
one
relates fundamental QCD parameters, such as quark masses,
renormalization
scales, vacuum condensates, etc., to hadronic parameters such as
particle
masses, widths, couplings, etc.. The convergence of the Hilbert
transform,
Eq.(7), may be improved by considering other integral kernels. This
leads to
other versions of QCD sum rules, such as the Laplace transform, FESR,
etc..

After this introduction we address the criticism raised in
\cite{ref11}. The
states $\mid n>$ entering Eq.(2) can be any complete set of states,
e.g.
hadronic states, quark-gluon basis, etc.. Eletsky and Ioffe claim that
below the critical temperature $T_{c}$ the suitable set of states is
the
hadronic set but not the quark-gluon basis. In support of this, they
argue
that for $T < T_{c}$ the original particles forming the heat bath,
being
probed by the external currents, are hadrons. They go on to say that
a summation over the quark-gluon basis in Eq.(2) would require taking
into
account the full range of their interaction, but that
no account of confinement was given  in \cite{ref1} - \cite{ref3}.

First
and foremost, we believe that the quark-gluon basis is indeed the
appropriate
basis to be used in QCD sum rule applications. If this were not the
case,
it would mean that the fruitful notion of duality would abruptly
loose meaning
as soon as the temperature is raised from T=0 to some arbitrary small
value of T. This would be a rather bizarre scenario.
According to the QCD sum rule philosophy, at T=0 one
calculates the theoretical left hand side of Eq.(7) through the OPE,
Eq.(3),
i.e. one uses quark-gluon degrees of freedom, and duality relates
this QCD
part to a weighted average of the hadronic spectral function. The
latter
arises from using  hadronic degrees of freedom. At very
low temperatures the hadronic spectrum will change very little, and
the external current will still convert into quark-antiquark pairs.
Hence,
it is only reasonable to assume that nothing drastic will happen with
duality. At finite temperature, though, there is a new effect coming
into
play, i.e. there are contributions to the QCD and hadronic spectral
functions in the space-like region (as opposed to only the time-like
region
at T=0). However, these additional contributions vanish smoothly as T
approaches zero, i.e. they do not introduce any discontinuous
behaviour.

Besides, the fact that the heat bath is mainly composed of
hadrons at small T is not in contradiction with the use of the quark-
gluon
basis. For instance, quarks enter the QCD perturbative term through
loops,
and have any value of momentum. One should emphasize here that the
currents
in (1) are external objects, and hence need not be in thermal
equilibrium
with the heat bath which is being probed.

Next, contrary to the statement made in \cite{ref11},
confinement has indeed been taken into account in  \cite{ref1} -
\cite{ref5} in the standard way, i.e. through the non-vanishing vacuum
condensates in the OPE, Eq.(3). The thermal beahaviour of these
condensates
is a separate matter. We have used in \cite{ref3} the chiral
perturbation
theory estimates of the thermal quark and gluon condensates
\cite{ref8},
and in \cite{ref4} we used the results of \cite{ref5}. Recently
\cite{ref10},
the T-dependence of four-quark condensates was obtained using soft
pion
techniques  in conjunction with Eq.(2), where the summation
was  performed in the hadronic (pion) basis. In this instance we
do agree that the hadronic basis is the appropriate one. However,
this has nothing to do with QCD nor with duality .
It is only one of many theoretical
approximations to estimate the temperature dependence of the
condensates.

Finally, we wish to comment on an implicit claim of Eletsky and Ioffe
\cite{ref11}, which was stated more explicitly  earlier in
\cite{ref12}.
This is, that calculating Gibbs averages in the quark basis at small
T
implies dealing with soft on-shell quarks, which are usually referred
to as
condensates. Although the quarks are on-shell, they can have any value
of  momentum when circulating around loops. And, there is
no possibility of confusing perturbative contributions
of on-shell quarks with quark condensates. This
is obvious at T=0 if one computes the imaginary part of a current
correlator, e.g. to one loop order, where the quark-antiquark
intermediate
state {\bf is} on-mass-shell. For instance, the leading perturbative
contribution to the imaginary part of $\Pi_{0}(q,T=0)$ in the axial-
vector
channel, calculated
using the Cutkosky rules (on-mass shell quarks understood) gives:
$Im \Pi_{0} = 1/8\pi$. This term cannot possibly be confused with the
quark
condensate contribution in Eq.(5) which, first, is real and, second,
it
has a different $Q^{2}$  dependence. We argue next, that this is also
the case at finite temperature.

With $q^{2}=\omega^{2} - {\bf q}^{2}$, and in the rest frame of the
medium
(${\bf q} \rightarrow 0$), the imaginary part of $\Pi_{0} (\omega
,T)$ in
the axial-vector channel, to leading order in perturbative QCD, is
\cite{ref3}
\begin{equation}
\frac{1}{\pi} \; \mbox{Im} \; \Pi_{0}^{(+)} (\omega, {\bf q} =
0,T) = \frac{1}{8 \pi^{2}} \; v (\omega) [3 - v^{2} (\omega)]
\mbox{th} \left( \frac{\omega}{4T} \right) \; \theta (\omega^{2} - 4
m_{q}^{2}) \; ,
\end{equation}
in the time-like region, and
\begin{equation}
\frac{1}{\pi} \; \mbox{Im} \; \Pi_{0}^{(-)} (\omega, {\bf q} =
0,T) = \frac{1}{8 \pi^{2}} \; \delta (\omega^{2}) \int_{4
m_{q}^{2}}^{\infty} \;
d z^{2} v (z) [3 - v^{2} (z)] 2 n_{F} \left( \frac{z}{2T} \right) \; ,
\end{equation}
in the space-like region, where $v(x)= (1 - 4
m_{q}^{2}/x^{2})^{1/2}$, and
$n_{F}$ is the Fermi thermal factor. The non-perturbative
contributions to
the OPE involve the quark and gluon condensates, both of dimension
d=4, the
four-quark condensate of dimension d=6, etc..All these are real, and
exhibit a temperature dependence very different from that in Eqs. (8)-
(9).
For instance, the quark condensate at low temperatures is of the form
$<<\bar{q} q>> = <\bar{q} q> (1 - a T^{2}/f_{\pi}^{2})$, while the
gluon condensate is essentially independent of T  \cite{ref8}. In
view of
these differences, we see no possibility of confusing thermal quark
loop
contributions with thermal quark condensates.

Additional evidence can be found in the framework of the O(N) sigma
model
in the large N limit, and in the Schwinger model, both in two-
dimensions.
These models were used in the past \cite{ref13} in order to verify
the
validity of the OPE. Since these two models can be solved exactly,
one can
compare the results of the exact calculation with that from the OPE.
They
turn out, in fact, to be identical. We obtain below the thermal Green
functions in these two models, and show that the temperature
corrections to
the perturbative contributions cannot be shifted to the non-
perturbative
terms, which develop their own T-dependence (calculable within the
framework
of these models).

We consider first the O(N) sigma model in 1+1 dimensions which is
characterized by the Lagrangian
\begin{equation}
{\cal{L}} = \frac{1}{2} [\partial_{\mu} \; \sigma^{a} (x)] \;
[\partial_{\mu} \sigma^{a} (x)] \; ,
\end{equation}
where a = 1,...N and $\sigma^{a} \sigma^{a} = N/f$, with f being the
coupling
constant. In the large N limit this model can be solved exactly (for
details see \cite{ref13}), it is known to be asymptotically free, i.e.
\begin{equation}
f(\mu) = \frac{4 \pi}{\ln \; \mu^{2}/m^{2}}
\end{equation}
and in spite of the absence of mass parameters in Eq.(10), it exhibits
dynamical mass generation. In addition, in this model there are vacuum
condensates, e.g. to leading order in 1/N
\begin{equation}
\langle 0| \alpha |0 \rangle = \sqrt{N} \; m^{2} \; ,
\end{equation}
whereas all other condensates factorize
\begin{equation}
\langle 0| \alpha^{k} |0 \rangle = (\sqrt{N} \; m^{2})^{k} \; ,
\end{equation}
In Eqs.(12)-(13) the $\alpha$ field is: $\alpha = f (\partial_{\mu}
\sigma ^{a})^{2}/\sqrt{N}$. We are interested in the Green function
associated with the propagation of quanta of the $\alpha$ field.
At T=0, in Minkowski space, and at the one loop order, this Green
function
is given by \cite{ref13}
\begin{eqnarray}
\Gamma (Q) & = & - \int \frac{d^{2} k}{(2 \pi)^{2}} \; \frac{i}{k^{2}
-
m^{2}} \; \frac{i}{(Q - k)^{2} - m^{2}} \nonumber \\[.4cm]
           & = &  \frac{1}{2 \pi} \;
            \frac{1}{\sqrt{Q^{2}(Q^{2}-4 m^{2})}} \; \ln \;
                 \frac{\sqrt{Q^{2}} - \sqrt{Q^{2} - 4 m^{2}}}
                 {\sqrt{Q^{2}} + \sqrt{Q^{2} - 4 m^{2}}}
\end{eqnarray}
Expanding Eq.(14) at short distances leads to
\begin{equation}
\Gamma (Q) =  \frac{1}{2 \pi} \; \frac{\ln (m^{2}/Q^{2})}{Q^{2}} \;
\left( 1 + \frac{2}{\sqrt{N}} \; \frac{\langle 0| \alpha |0
\rangle}{Q^{2}}
+ \frac{6}{N} \; \frac{\langle 0| \alpha^{2} |0 \rangle}{Q^{4}} \; +
\; \cdots
\right) \; ,
\end{equation}
A separate calculation, based on the OPE of the current correlator
involving
the scalar current $J_{S} = f (\partial_{\mu} \sigma^{a})^{2}$, gives
exactly the same answer as Eq.(15) \cite{ref13}.

We have calculated the Green function Eq.(14) at finite temperature.
Its
imaginary part can be integrated analytically in closed form and is
\begin{eqnarray}
\mbox{Im} \; \Gamma (\omega, {\bf q} = 0, T) &=&
\frac{1}{2 \omega^{2}} \left[ 1 + 3 n_{B} \; (\omega/2T) \right]
\nonumber \\[.4cm]
& + & \frac{1}{2} \left[ \frac{2}{\sqrt{N}} \;
\frac{<< \alpha >>}{\omega^{4}} + \frac{6}{N} \;
\frac{<< \alpha^{2} >>}{\omega^{6}} \;  + \; \cdots \right] \; ,
\end{eqnarray}
where the first term above corresponds to the perturbative
contribution,
the second to the non-perturbative, and $n_{B}$ is the thermal Bose
factor.
Equation (16) is valid in the time-like region; the space-like region
counterpart vanishes in 2 dimensions.
Since the model is exactly solvable, the thermal behaviour of the
vacuum
condensates can also be calculated, viz.
\begin{equation}
<< \alpha >> = < \alpha > \left[ 1 + 3 n_{B} (\omega/2T)
\right]
\end{equation}
In this case the vacuum condensates contribute to the imaginary part,
and
as Eq.(16) shows, the thermal dependence of the
perturbative piece cannot be absorbed into the condensates. Hence, no
confusion should arise. We have not discussed above the
effects of renormalization \cite{ref13}, as they are not essential
to our argument.

Finally, we consider the Schwinger model in 1+1 dimensions, with the
Lagrangian
\begin{equation}
{\cal{L}} = - \frac{1}{4} F_{\mu \nu} \;
F_{\mu \nu} + \bar{\psi} i \;
\gamma_{\mu} \; {\cal{D}}_{\mu} \; \psi
\end{equation}
where ${\cal D}_{\mu} = i \partial_{\mu} + e A_{\mu}$. This model has
been
solved exactly \cite{ref14}, and in the framework of the OPE
\cite{ref13}.
The short distance expansion of the exact solution coincides with that
from the OPE, as shown in \cite{ref13}.
Here, we are interested in the two-point functions
\begin{equation}
\Pi_{++} (x) = \langle 0 |T \{j^{+} (x) j^{+} (0) \}|0 \rangle
\end{equation}
\begin{equation}
\Pi_{+-} (x) = \langle 0 |T \{j^{+} (x) j^{-} (0) \}|0 \rangle
\end{equation}
where the scalar currents are
\begin{equation}
j^{+} = \bar{\psi}_{R} \; \psi_{L} \; \; \; \; \; ,
\; \; \; \; j^{-} = \bar{\psi}_{L} \; \psi_{R}
\end{equation}
with $\psi_{L,R} = (1 \pm \gamma_{5}) \psi/2$. The function $\Pi_{++}
(Q)$
vanishes identically in perturbation theory, and the leading non-
perturbative
contribution involves a four-fermion vacuum condensate. At T=0
one finds \cite{ref13}
\begin{equation}
\Pi_{++} (Q, T=0) = \frac{16 \; e^{2}}{Q^{6}}
\langle 0 | \bar{\psi}_{R} \; \partial_{\mu} \; \psi_{L} \;
\bar{\psi}_{R} \;
\partial_{\mu} \; \psi_{L} | 0 \rangle
\end{equation}
On the other hand, the function $\Pi_{+-} (Q)$ is purely
perturbative. At
T=0 it is given by \cite{ref13}
\begin{equation}
\Pi_{+-} (Q, T=0) = \frac{1}{4 \pi} \; \left( \ln
\frac{M^{2}}{Q^{2}} + \frac{m^{2}}{Q^{2}} + \cdots \right) \;,
\end{equation}
where M is an ultraviolet cutoff.
We have calculated the thermal behaviour of these  current
correlators and
obtain, e.g. for their imaginary parts in the time-like region (there
is
no space-like contribution in 2 dimensions)
\begin{equation}
\mbox{Im} \; \Pi_{++} (\omega, {\bf q} = 0,T) = 0
\end{equation}
\begin{equation}
\mbox{Im} \; \Pi_{+-} (\omega, {\bf q} = 0,T) = \frac{1}{4} \;
\left[ 1 - 2n_{F} (\omega/2T) \right]
\end{equation}
Hence, the choice of the fermion basis in the Gibbs average of current
correlators does not imply confusing these fermions with condensates.
The (perturbative) fermion loop terms and the
(non-perturbative) vacuum condensates develop their own temperature
dependence. In this particular example, non-perturbative terms are
totally
absent in $Im \Pi_{+-}$, and they do not contribute to $Im \Pi_{++}$
because they are real. This peculiarity makes the argument particulary
transparent.

It should be mentioned, in closing, that there are some unresolved
problems in finite temperature QCD sum rules, e.g. when one tries
to use them to extract the thermal behaviour of the vacuum condensates
\cite{ref15}. However, these problems are unrelated to the arguments
given by Eletsky and Ioffe \cite{ref11}.

\subsection*{Acknowledgements}
The authors are indebted to Ninoslav Bralic and Olivier Espinosa for
enlightening discussions.
The work of (CAD) has been supported in part by the Foundation for
Research Development and Fundaci\'{o}n Andes, and that of (ML)
by FONDECYT 0751/92. This
work has been performed in the framework of the FRD-CONICYT Scientific
Cooperation Program. (CAD) wishes to thank the Facultad de Fisica,
Pontificia Universidad Cat\'{o}lica de Chile for their kind
hospitality.

\end{document}